\documentclass[apj,iop]{emulateapj}
 
\usepackage{graphicx} 
\usepackage{apjfonts} 

\shorttitle{Internetwork flux appearance and disappearance rates} 

\shortauthors{Go\v{s}i\'{c} et al.}

\begin{document}

\title{The solar internetwork. II. Magnetic flux appearance and disappearance
  rates}

\author{M.~Go\v{s}i\'{c}$^1$}
\author{L.~R.~Bellot Rubio$^1$}
\author{J.~C.~del Toro Iniesta$^1$}
\author{D.~Orozco Su\'arez$^2$}
\author{Y.~Katsukawa$^3$}

\affil{$^1$ Instituto de Astrof\'{\i}sica de Andaluc\'{\i}a (CSIC),
  Apdo.\ 3004, E-18080 Granada, Spain; mgosic@iaa.es}
\affil{$^2$Instituto de Astrof\'{\i}sica de Canarias, E-38205 La Laguna, 
Tenerife, Spain}
\affil{$^3$ National Astronomical Observatory of Japan, 2-21-1 Osawa, 
Mitaka, Tokyo 181-8588, Japan}

\begin{abstract}
  Small-scale internetwork magnetic fields are important ingredients
  of the quiet Sun. In this paper we analyze how they appear and
  disappear on the solar surface. Using high resolution \textit{Hinode}
  magnetograms, we follow the evolution of individual magnetic
  elements in the interior of two supergranular cells at the disk
  center. From up to 38 hr of continuous measurements, we show that
  magnetic flux appears in internetwork regions at a rate of
  $120\pm3$~Mx~cm$^{-2}$~day$^{-1}$ ($3.7 \pm 0.4 \times
  10^{24}$~Mx~day$^{-1}$ over the entire solar surface).  Flux
  disappears from the internetwork at a rate of
  $125 \pm 6$~Mx~cm$^{-2}$~day$^{-1}$ ($3.9\pm 0.5 \times 10^{24}$~Mx~day$^{-1}$)
  through fading of magnetic elements, cancellation between
  opposite-polarity features, and interactions with network patches,
  which converts internetwork elements into network features. Most of
  the flux is lost through fading and interactions with the network,
  at nearly the same rate of about 50~Mx~cm$^{-2}$~day$^{-1}$. Our
  results demonstrate that the sources and sinks of internetwork
  magnetic flux are well balanced. Using the instantaneous flux
  appearance and disappearance rates, we successfully reproduce the
  time evolution of the total unsigned flux in the two supergranular
  cells.
\end{abstract}


\section{Introduction}

Internetwork (IN) magnetic fields are observed to fill the interior of
supergranular cells, enclosed by the photospheric network (NE). In
Paper~I of this series \citep{Gosic2014} we showed that IN regions
harbor some $10^{23}$~Mx over the entire solar surface, which accounts
for 15\% of the total quiet Sun (NE$+$IN) flux. We also demonstrated
that IN fields represent the main source of flux for the NE, capable
of supplying as much flux as it contains in only about 10~hr. This
renders the IN an essential contributor to the flux budget of the
solar photosphere---and probably also to its energy budget
\citep{2004Natur.430..326T}.

\begin{figure*}[t]
\begin{center}
\resizebox{1\hsize}{!}{\includegraphics[]{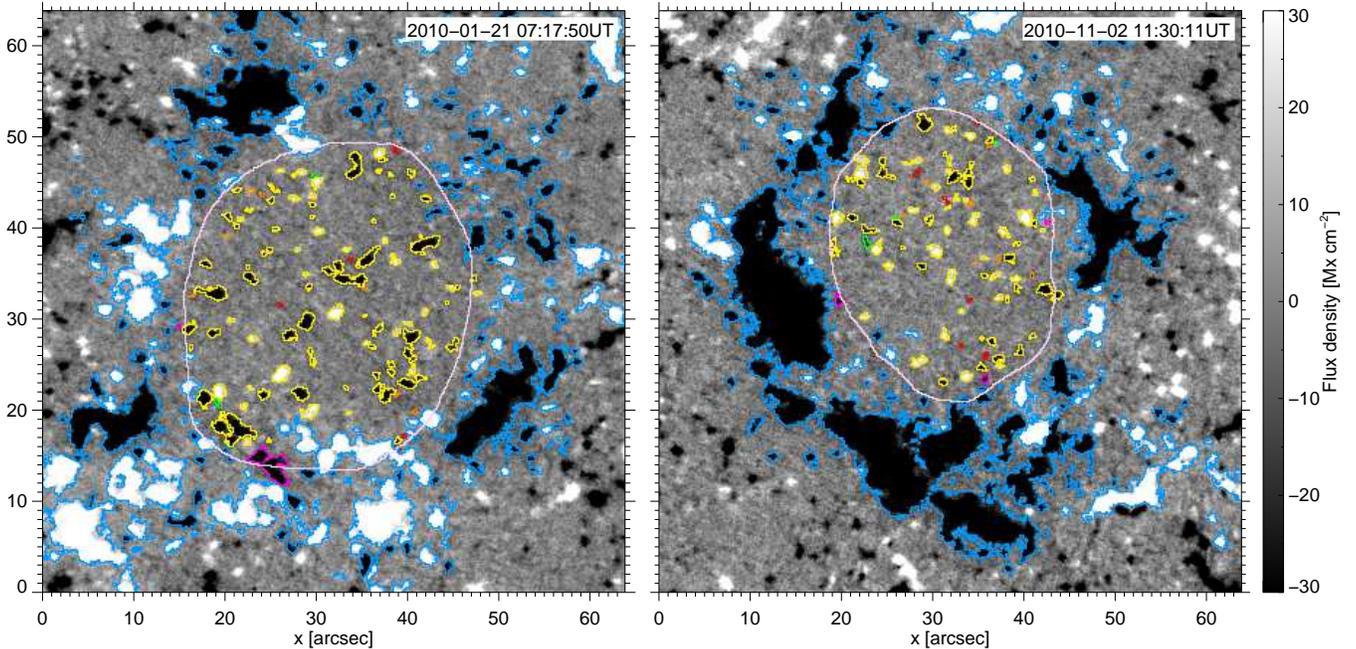}}
\end{center}
\vspace*{0em}
\caption{Individual supergranular cells observed in data sets 1 and 2
  (left and right, respectively). The cell interiors are outlined by
  pink contours and the surrounding NE flux features are marked with
  blue contours. Red contours show IN elements that appear in situ in
  these frames. Magnetic flux disappears from the cell interiors
  when elements fade (orange), cancel out (green), or enter the NE and
  interact with NE features (purple). IN patches that do not undergo
  any of those processes in the displayed frames are indicated with
  yellow contours.\newline
  {\em An animation of the right panel is available.}}
\label{fig1}
\end{figure*}

IN regions show an enormous flux appearance rate. The values quoted in
the literature range from $10^{24}$~Mx~day$^{-1}$ \citep{Zirin1987} to
$3.8\times10^{26}$~Mx~day$^{-1}$ \citep{Zhou2013}. All of them
are significantly larger than the $6 \times 10^{21}$~Mx~day$^{-1}$
brought to the surface by active regions during the maximum of the
solar cycle \citep{1994SoPh..150....1S}. However, the published rates
vary by more than 2 orders of magnitude, indicating that they are not
well determined yet. Much of the difference is probably due to the
methods used to derive them. For example, some authors rely on
automatic tracking algorithms while others prefer a manual detection
of magnetic elements. Both approaches are affected by uncertainties
and/or subjectivity \cite[e.g.,][]{DeForest}. Differences in the
spatial resolution, noise level, cadence, and duration of the
observations may also lead to different results. Another source of
disagreement is the fact that many estimates are based on fitting the
observed flux distributions with power laws, rather than on following
the evolution of individual elements.  This method assumes constant
appearance rates, but in reality we do not know if the flux appears at a 
constant rate or there are spatial and temporal variations across the solar 
surface.

To understand the flux balance of IN regions, we have to evaluate the
processes by which flux is injected into and removed from the interior
of supergranular cells. The IN gains flux through in-situ appearance
of magnetic elements. Sometimes bipolar structures can be clearly
identified, but most elements seem to appear as unipolar patches. On
the other hand, flux is removed from the cell interior by fading
(disappearance of features without obvious interactions with other
elements) and cancellation (total or partial disappearance of features
of opposite polarity as they come into close contact).  Fading has
been reported to be the dominant process, accounting for 83\% of the
flux removal in the quiet Sun \citep{Lamb2013}. These authors
determined a flux disappearance rate of
$2.4\times10^{26}$~Mx~day$^{-1}$ over the entire solar surface, but
they did not distinguish between NE and IN regions.

Another important sink of IN flux that has been neglected until now 
is conversion of IN features into NE elements. In Paper~I we made the first 
steps to quantify this process, showing that 40\% of the IN flux eventually 
interacts with NE elements and disappears from the supergranular cell. Flux 
transfer from the IN to the NE is therefore an essential ingredient to 
understand the flux balance of quiet Sun regions.

In this paper we determine the rates of the four processes mentioned
above, namely in-situ appearance, fading, cancellation, and flux
transfer to the NE. We focus on two well-defined supergranular cells
and compute the instantaneous fluxes that appear and disappear in
their interiors. The analysis is based on long-duration, high-cadence
magnetogram sequences taken with the Narrowband Filter Imager
\cite[NFI;][]{Tsuneta} on board \textit{Hinode} \citep{2007SoPh..243....3K}.
These data are perfectly suited to study the highly dynamical IN
fields on temporal scales from minutes to days.  The evolution of IN
flux elements is followed using an automatic feature tracking
algorithm and a new code developed to accurately resolve interactions
between flux patches (Section~\ref{method}).  In this way we are able
to compute, for the first time, the flux appearance and disappearance
rates in individual supergranular cells and their variations with time
(Section~\ref{results}).

\section{Observations and data processing}
\label{sect2}

The data used in this paper consist of two temporal sequences acquired
with the \textit{Hinode} NFI on 2010 January 20--21 and 2010 November 2--3
(data sets 1 and 2, respectively). They belong to \textit{Hinode} Operation
Plan 151. The observations have been described in detail in Paper~I,
so here we only summarize their main features. Magnetograms and
Dopplergrams were constructed from Stokes \textit{I} and \textit{V} filtergrams 
taken at $\pm 16$~pm from the center of the magnetically sensitive \ion{Na}{1}
589.6~nm line. We observed both line wings to make the magnetograms as
independent of Doppler shifts as possible. In addition, we pushed the
sensitivity of the observations to a limit by operating the NFI in
shutterless mode.  This allowed us to reach a total exposure time of
6.4~s per magnetogram, resulting in a noise level of 6~Mx~cm$^{-2}$.
The noise was further reduced to 4~Mx~cm$^{-2}$ through application of
a $3\times3$ Gaussian-type spatial smoothing kernel. Thanks to these
choices, our magnetogram sequences are among the most sensitive ever
obtained in the quiet Sun with a filter instrument.

To study the evolution of the solar IN we made continuous measurements
for 20~hr in the case of data set~1 and 38~hr in the case of data
set~2. The sequences show a few gaps due to telemetry problems, but
their duration is often short (of the order of minutes). The achieved
cadences---60 and 90~s---are ideal for automatic tracking of magnetic
features. 

The observations cover large areas of the quiet Sun at disk center
($82\arcsec \times 113\arcsec$ and $80\arcsec \times 74\arcsec$,
respectively). Solar rotation was compensated, making it possible to
monitor individual supergranular cells for long periods of time. Here
we focus on two cells that were visible at the center of the field of
view during the entire sequences (Figure \ref{fig1}). They show an
average total (NE$+$IN) unsigned flux of $2.4\times10^{20}$~Mx and
$3.6\times10^{20}$~Mx, respectively. We observed their evolution from
the early formation phases until they fully developed into mature
cells. Their effective radius increased from $\sim$9.7 to $\sim$13~Mm
during the process. Neither of them underwent mergings or
fragmentations. Thanks to the longer duration of the second data set,
we witness how the supergranule disperses with time and slowly loses
its form, although it is still visible by the end of the observations.
This cell is strongly unipolar and shows a net flux of
$-2.3\times10^{20}$~Mx (enhanced NE). By contrast, the supergranular
cell of data set 1 is in almost perfect polarity balance with a net
flux of $-2\times10^{18}$~Mx (quiet NE).

\begin{figure*}[t]
\begin{center}
\resizebox{.96\textwidth}{!}{\includegraphics{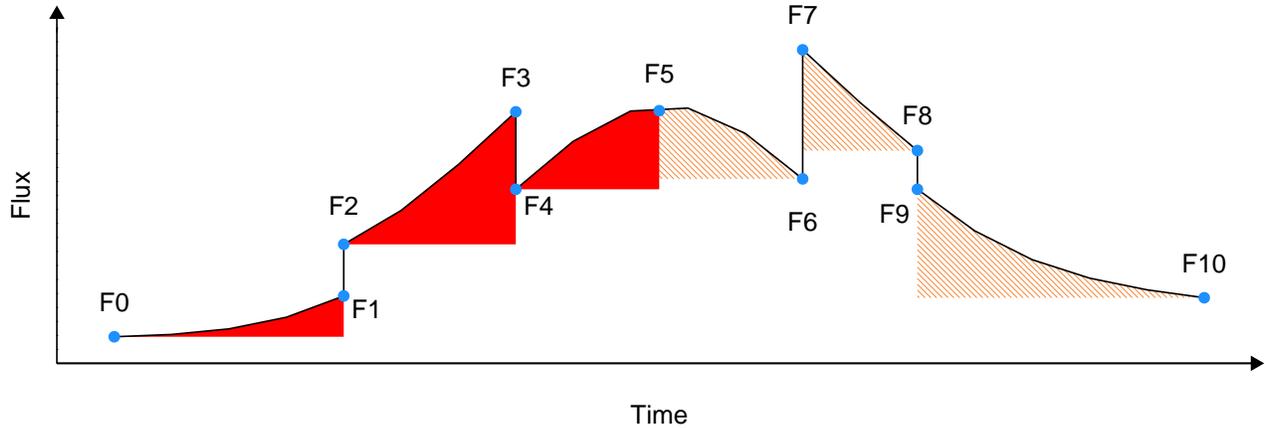}}
\end{center}
\vspace*{-1em}
\caption{Sketch of the flux evolution of typical IN magnetic elements.
  The red shaded areas represent the flux gained by an element during
  its lifetime. The orange shaded areas represent the flux lost by
  fading until the element disappears (by either in-situ fading,
  cancellation, or merging with a stronger flux patch).  Blue dots
  mark the moments of appearance, disappearance, and interactions.}
\label{scheme}
\end{figure*}

\section{Method}
\label{method}

Our goal is to determine the instantaneous flux appearance and
disappearance rates in the two supergranular cells described above.
This requires us to detect and track all the individual magnetic
elements visible in their interiors. As explained in Paper~I, the cell
boundaries are known precisely from horizontal flow divergence maps
derived from the available Dopplergrams through Local Correlation
Tracking \citep{November}, and turn out to be pretty well outlined by
the strong patches of the NE.

We use the YAFTA code \citep[]{WelschLongcope} and the clumping method
to automatically identify and track individual elements in the
magnetogram sequences.  Only features that have at least 4 pixels with
flux densities above 12~Mx~cm$^{-2}$ (three times the noise level) and
live for 2 or more frames are taken into consideration. We have
developed a new code to solve some of the problems that lead to
misidentifications and/or incorrect labeling of the magnetic elements
(Go\v{s}i\'c et al., in preparation). By use of this code we can
follow individual elements from birth to death, which is necessary to
derive reliable appearance and disappearance rates. The quality of the
final tracking can be assessed from the animation accompanying
Figure~\ref{fig1}.

Supergranular cells gain flux through in-situ appearance of magnetic
elements and their subsequent evolution. The flux contributed by an
element to the IN is not only its initial flux, but also any intrinsic
flux increase it may experience with time. We refer to the sum of both
contributions as total appeared flux. Flux disappears from the cell
when IN elements fade inside the supergranule, cancel totally or
partially with opposite-polarity IN elements, or leave the
supergranule and interact with NE elements (a process we refer to as
flux transfer to the NE).  Below we explain how the rates of these
processes are calculated and describe the problems solved by our code
in each case.

\subsection{Flux sources}

The flux that appears in situ in a given frame is computed considering
all IN patches that become visible for the first time in that frame.
Identifying those elements is easy because YAFTA tags them with new
labels. If the elements do not interact with other patches, their
maximum flux is taken to be the flux they bring to the solar surface.
IN elements tend to increase in flux upon appearance, so the maximum
flux they reach is usually larger than the initial one.  When
interactions occur, however, the maximum flux is not a good indicator
of the flux appearing on the surface, because the elements may simply
grow through mergings with like-polarity patches. The flux gained in
this way does not represent new flux and should not be counted.
In addition, elements that appear in situ can fragment a few frames
later and the fragments themselves may continue to gain flux.  Such a
flux increase needs to be added to the flux appearance rate. Fragments
are tagged by YAFTA as children of their parents, so they are easy to
detect.

Thus, one has to be careful with interactions to derive reliable flux
appearance rates. In practice, we go through each of the magnetic
elements assigned a new label by YAFTA in a given frame---either
because they appeared in situ or because they fragmented from an
existing element.  We examine how they gain flux over time, correct
for mergings and fragmentations, and sum all the contributions to
get the total flux they bring to the surface.  The process is
illustrated schematically in Figure~\ref{scheme}. The solid line
represents the flux of an IN element during its evolution.  The moments
of appearance, disappearance, and interactions are marked with dots.
Each dot has a value from $F0$ to $F10$, which corresponds to the flux
of the element at those particular times. $F2$ and $F7$
coincide with sudden flux increases due to mergings with weaker elements.
The maximum observed flux is $F7$, but this value is the result of one
such merging.  $F5$ would be the maximum flux in the absence of
interactions.  There are two fragmentations which decrease the flux
from $F3$ to $F4$ and from $F8$ to $F9$, respectively.

We calculate the total flux the element brings to the supergranular
cell by adding together the initial flux $F_{\rm init}$ and the flux
increases from one checkpoint to the next (blue dots), i.e.,
\begin{equation}
F_{\rm app}=F_{\rm init} + (F1-F0)+(F3-F2)+(F5-F4).
\label{equation1}
\end{equation}
If the element appeared in situ, we take $F_{\rm init}=F0$. If it is
a fragment of an existing patch, then $F_{\rm init}=0$.
The flux $F_{\rm init}$ is assigned to the moment when the element was
first detected (also for fragments). The other contributions 
in Equation \ref{equation1} are evenly distributed over their
corresponding time intervals.

By adding the flux carried by the patches that appear in situ and
through fragmentation we obtain the total flux appearance rate. For
completeness, we also compute the appearance rate using only the
initial flux of the features that appear in situ. We do this because
the initial flux is a well-defined quantity that does not depend on
interactions between patches. This rate represents the flux that
discrete magnetic elements bring to the solar IN right at the moment
of appearance, before they experience any perturbation.

It should be clear by now that a very accurate tracking of magnetic
elements is mandatory to obtain reliable flux appearance rates. In
particular, false detections of disappearances followed by appearances
of the same element must be avoided. This problem occurs frequently
due to fluctuations in the signal of the elements. They may exhibit
flux densities below the detection threshold during a few frames, only
to return to their previous state after such episodes.  YAFTA
interprets this evolution as the disappearance of the element by
fading and the subsequent appearance of a new like-polarity patch at
the same location, artificially increasing the appearance and
disappearance rates.  Our code corrects for this problem in the
following way.

For each element disappearing in situ, we check whether a new
like-polarity flux structure appears nearby in the next three frames.
In that case, the new patch is considered to be the evolution of the
previously disappeared element if three conditions are met.  First,
the newly appeared element has to have at least 30\% of its area
inside a circle centered on the flux-weighted center of the
disappeared element. The circle radius is 2 pixels (230~km) in the
frame after the disappearance occurred, and 4 pixels afterward.
Second, the size and total flux of the two patches cannot differ by
more than a factor of three.  Third, if any of the two elements is
visible in only one frame, then they have to overlap by more than
70\%. This correction decreases both the flux appearance and
disappearance rates by 10\%, compared with the original YAFTA results.

Another problem fixed by our code is in-situ appearance of IN elements
followed by merging with existing NE patches in the same frame. YAFTA
does not detect these IN features and therefore their flux is not
included in the flux appearance rates. We identify such elements by
comparing the shapes of the NE elements that are inside the cell with
the shapes they had in the previous frame, looking for large size
changes. When we find 16 or more contiguous pixels in the
non-overlapping area, they are taken to be a newly appeared IN element
which merged with the NE patch right away. In our data sets, such a
correction increases the appearance rates by approximately 5\%. Yet,
this is only a lower limit because structures smaller than 16 pixels
are not counted in (to avoid errors induced by the intrinsic shape
variation of NE patches from frame to frame).

\subsection{Flux sinks}

The simplest process of flux disappearance from supergranules is
in-situ fading, whereby IN magnetic elements disappear in a given
frame without interacting with other features in their vicinity.
YAFTA tags fading events as in-situ disappearances, so their detection
is relatively straightforward (but see below). The flux they remove
from the IN is calculated in a similar way as the flux that appears
on the solar surface. If the element does not undergo
interactions, the total flux lost by fading is the maximum flux it
attains during its lifetime. When interactions occur, we determine
how the flux decreases from one interaction to the next.  In the
example of Figure~\ref{scheme}, the total flux removed by fading is
\begin{equation}
F_{\rm fading}=(F5-F6)+(F7-F8)+(F9-F10) + F_{\rm final}.
\end{equation} 
$F_{\rm final} = F10$ if the element disappears completely by fading.
If the patch loses its label at $F10$ because of a merging with a
stronger feature (either from the IN or from the NE), then $F_{\rm
  final} = 0$ to avoid counting the flux it had right before the
merging. Thus, even though mergings do not remove flux from the
photosphere, we still calculate the flux lost by fading before the
elements merge with stronger features. The flux $F_{\rm final}$ is
assigned to the frame where the element is detected for the last 
time, and the remaining terms to their respective intervals.

The second mechanism capable of removing flux from the IN is
cancellation of opposite-polarity patches. It turns out that YAFTA
does not identify this as a process different from in-situ
disappearance.  To detect cancellations, we use the YAFTA output and
look for IN elements that disappear at most 2 pixels away from an
opposite-polarity IN patch. The flux lost by the supergranular cell
through this process is equal to the flux the two canceling features
had at the beginning of the cancellation. If the magnetic elements
merge with other patches or fragment during the process, we keep track
of the changes and revise the total canceled flux accordingly. In
partial cancellations with one surviving feature, the canceled flux is
taken to be twice the flux of the feature that disappeared completely.
Partial cancellations where neither of the elements disappear are very
difficult to detect, so we do not account for them. As a consequence,
our flux cancellation rates may be smaller than the actual ones.
However, the total disappearance rate is correct, because the flux
removed in those cases is counted as fading flux. For all the magnetic
elements that cancel out, we check if they lose flux before the
cancellation starts. Any observed flux drop is ascribed to fading.

\begin{figure*}
\begin{center}
\resizebox{\hsize}{!}{\includegraphics[bb=20 0 895 453]{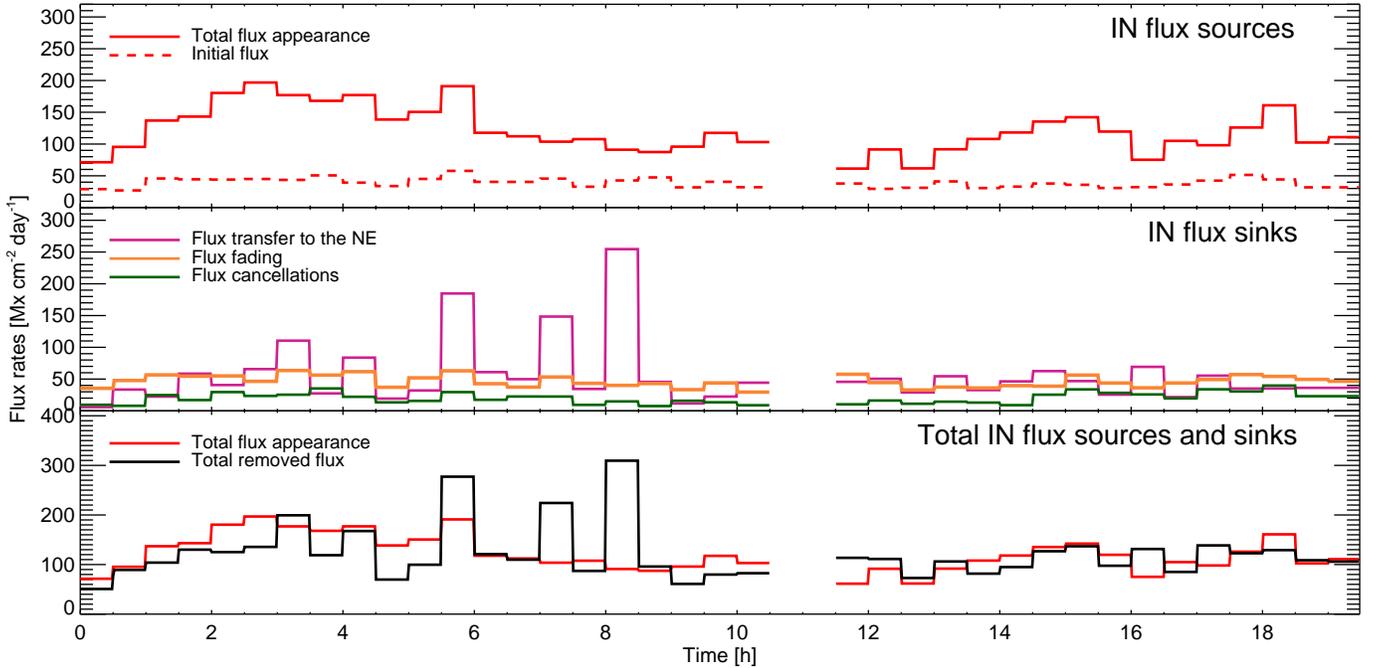}}
\end{center}
\vspace*{-1em}
\caption{Rates at which the central supergranule of data set~1 gains
  and loses magnetic flux. The data are binned in 30 minute intervals.
  Top panel: the source of IN flux is in-situ appearance of
  magnetic features and their subsequent evolution (red solid curve).
  The red dashed line shows the initial flux of the elements that
  appear in situ. Middle panel: flux is removed from the cell by
  three mechanisms, namely interaction of IN patches with NE elements
  (flux transfer to the NE; purple), in-situ disappearance of magnetic
  elements (orange), and cancellation of IN features (green). Bottom panel: 
total rates at which flux is accumulated (red solid line) and removed (black 
solid line) from the cell. The latter is defined as the sum of the flux sinks 
displayed in the middle panel.}
\label{fig2}
\end{figure*}

The third mechanism of flux removal from the IN is transfer to the NE.
Many IN elements are sufficiently persistent to drift toward the
supergranular border \citep[e.g.,][]{Orozco2012}, where they merge or
cancel with NE patches. In both cases, the IN loses flux. This complex
process has been characterized in Paper~I. The basic idea is to follow
the evolution of all IN flux structures to determine if they interact
with NE patches.  The flux lost by the IN is taken to be the flux of
the IN element in the frame before the merging or cancellation
occurred. More details about these calculations are provided in
Paper~I. 

Finally, we note that the IN flux our code detects to appear in situ and
immediately merge with NE features is counted also as IN flux transferred
to the NE, which increases the total flux disappearance rate by almost
5\% compared with the plain YAFTA results.

\subsection{Uncertainties}
\label{uncertainties_section}
The main source of error in the YAFTA tracking is the
misidentification of magnetic elements that fade and reappear again.
As explained above, the method we use to correct for this problem
depends on a number of (free) parameters such as circle radius,
minimum area overlap, and maximum flux and size variations. These
parameters have been adjusted manually to yield the best possible
results, but of course they introduce uncertainties in the flux
appearance and disappearance rates. In what follows we estimate them.

We use two limiting cases. The first case represents the plain YAFTA
results.  Here, many magnetic features are erroneously identified as
new elements when their fluxes drop below and later rise 
above the $3\sigma$ level.

The other limiting case is computed using only one parameter, namely
the radius of the circle in which we look for reappearing elements.
If a new magnetic feature shows up inside or touches the border of a
circle of radius 4 pixels, then that feature is considered to be the
continuation of the previously disappeared element, regardless of its
flux or size. This criterion is too lax, producing bad identifications
of many elements that are detected as reoccurring patches. The
appearance and disappearance rates in this extreme case are 23\% and
19\% lower than those from the non-corrected YAFTA output,
respectively. 

The combination of parameters we actually use in the analysis leads to
more strict criteria for the identification of these elements.  Thus,
our final appearance and disappearance rates are 10\% lower than those
obtained with the plain YAFTA tracking, and 13\%-9\% larger than
those resulting from the second limiting case. Therefore, an upper
limit for the uncertainty caused by the choice of parameters is
$\pm13$\% (appearance rate) and $\pm10\%$ (disappearance rate).

\section{Results}
\label{results}

Figure~\ref{fig2} displays the rates at which the IN gains and loses
flux in the supergranular cell of data set 1.  The data points are
binned in 30 minute intervals.

The top panel shows the total flux appearance rate as a function of
time (solid line). On average, newly appeared features bring
117~Mx~cm$^{-2}$~day$^{-1}$ to the surface.  The appearance rate
increases when strong magnetic elements pop up in the interior of the
cell in the form of clusters. This is what happened, for example, between 1
and 6~hr. During that interval, the instantaneous appearance rate 
nearly doubled, reaching $\sim$200~Mx~cm$^{-2}$~day$^{-1}$. The flux
appearance rate computed using the initial flux of the elements is
38~Mx~cm$^{-2}$~day$^{-1}$, with very little fluctuations (dashed
line).

The middle panel of Figure~\ref{fig2} shows the flux removed from the
cell by interactions with NE patches (purple curve), fading (orange
curve), and cancellations (green curve).  Transfer of magnetic
elements from the IN to the NE turns out to be a very important
process of flux disappearance from the supergranule, at an average
rate of 53~Mx~cm$^{-2}$~day$^{-1}$.  This process shows large temporal
variations in the selected cell. The peaks between 6 and 8~hr, for
example, were produced by strong patches leaving the IN and merging
with the NE. In addition to this mechanism, flux disappears from the
supergranule through fading, at a rate of 46~Mx~cm$^{-2}$~day$^{-1}$,
and through cancellations, at a rate of 20~Mx~cm$^{-2}$~day$^{-1}$.
Thus, fading and transfer to the NE are equally important sinks of
flux for this cell, with cancellations playing a secondary role.

\begin{figure*}
\begin{center}
\resizebox{\hsize}{!}{\includegraphics[bb=20 0 895 453]{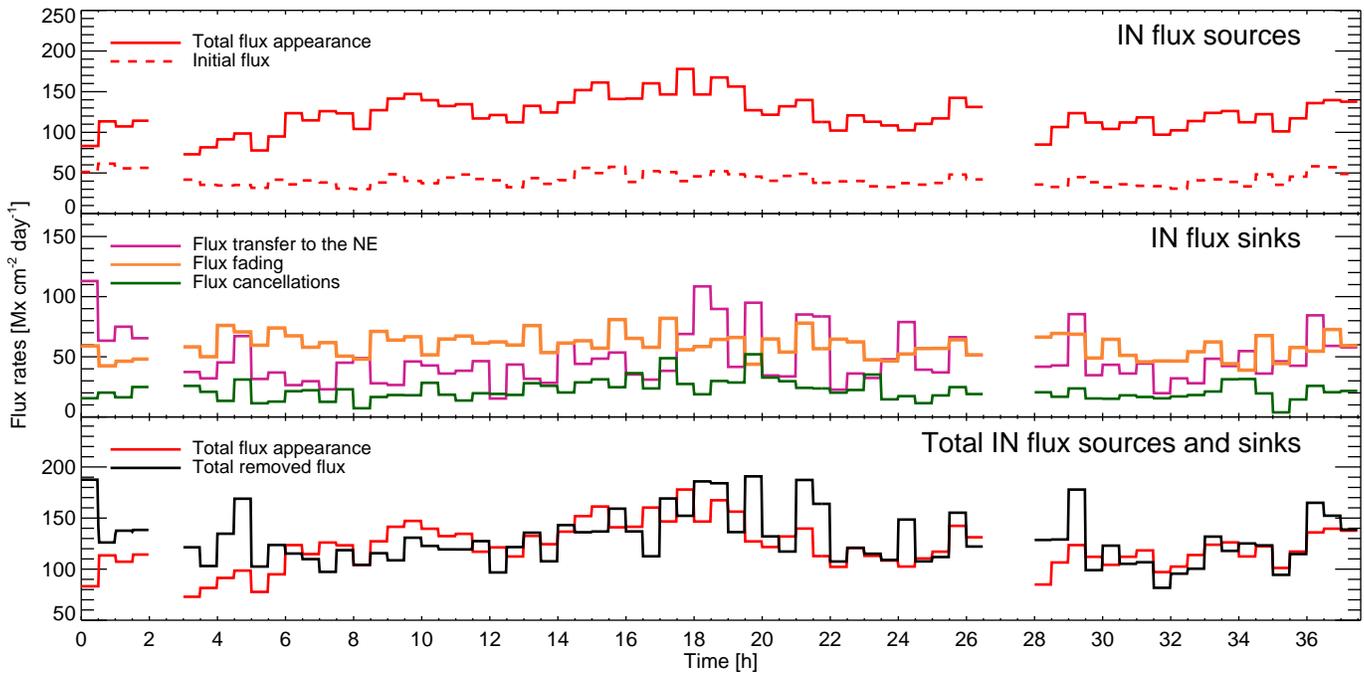}}
\end{center}
\vspace*{-1em}
\caption{Same as Figure \ref{fig2}, for the central supergranule of
  data set~2.}
\label{fig3}
\end{figure*}

The sources and sinks of magnetic flux are shown in the bottom panel
of Figure~\ref{fig2} with red and black lines, respectively. As can be
seen, the flux appearance and removal rates are similar, but their
peaks do not coincide exactly. The reason is that magnetic elements
grow in flux upon appearance, reach their maximum, and disappear at a later
time, depending on their intrinsic evolution and the interactions they
undergo.

Figure~\ref{fig3} summarizes the results for the supergranular cell of
data set~2. We observe an average total flux appearance rate of
122~Mx~cm$^{-2}$~day$^{-1}$. Fluctuations occur when clusters of
magnetic patches appear in the cell interior. They can be seen at
around 10 and 18~hr, for example. By contrast, the initial flux
appearance rate is very stable at 41~Mx~cm$^{-2}$~day$^{-1}$.

Also in this case, transfer of IN flux to the NE and fading are the
main flux removal mechanisms (second panel of Figure~\ref{fig3}).  The
rates at which the cell loses flux through interactions with the NE is
47~Mx~cm$^{-2}$~day$^{-1}$. Fading is slightly larger, accounting for
59~Mx~cm$^{-2}$~day$^{-1}$, while cancellations proceed at a rate of
25~Mx~cm$^{-2}$~day$^{-1}$. In general, IN patches that convert into
NE elements are small, but occasionally we detect strong IN features
carrying much flux away from the cell interior. These events are
easily distinguishable at, e.g., the beginning of the time sequence or
18~hr. The cancellation rates are smaller but more stable than the
flux transfer to the NE.

Just like in data set 1, the supergranular cell of data set~2 gains
and loses flux at nearly the same rate (Figure~\ref{fig3}, bottom panel).

\begin{deluxetable}{lrrr}
  \tablecolumns{4} \tablewidth{\columnwidth} 
  \tablecaption{IN flux appearance and disappearance rates (Mx~cm$^{-2}$~day$^{-1}$)} 
  \tablehead{ \colhead{} & \colhead{Data Set 1} & \colhead{Data Set 2} & 
  \colhead{Mean}} 
  \startdata
  Appearance & & & \\
  \hspace{1em} In-situ & 117 & 122 & $120 \pm 3$ \\
  Disappearance & & & \\
  \hspace{1em} Fading & 46 & 59 & $53 \pm 7$ \\
  \hspace{1em} Cancellation & 20 & 25 & $23 \pm 3$ \\
  \hspace{1em} Transfer to NE & 53 & 47 & $50\pm 3$ \\
  \hspace{1em} Total & 119 & 131 & $125 \pm 6$ \\
  Disappearance/Appearance & 1.02 & 1.07 & $1.04 \pm 0.03$
\enddata
\label{table1}
\end{deluxetable}

\section{Discussion}
\label{discussion}

Our results convey information on and have implications for the flux
balance of IN regions, the magnetic flux history of supergranular
cells, and the mechanisms responsible for the appearance of flux on
the solar surface. We discuss these implications in the following.

\begin{figure*}
\begin{center}
\resizebox{.97\hsize}{!}{\includegraphics[]{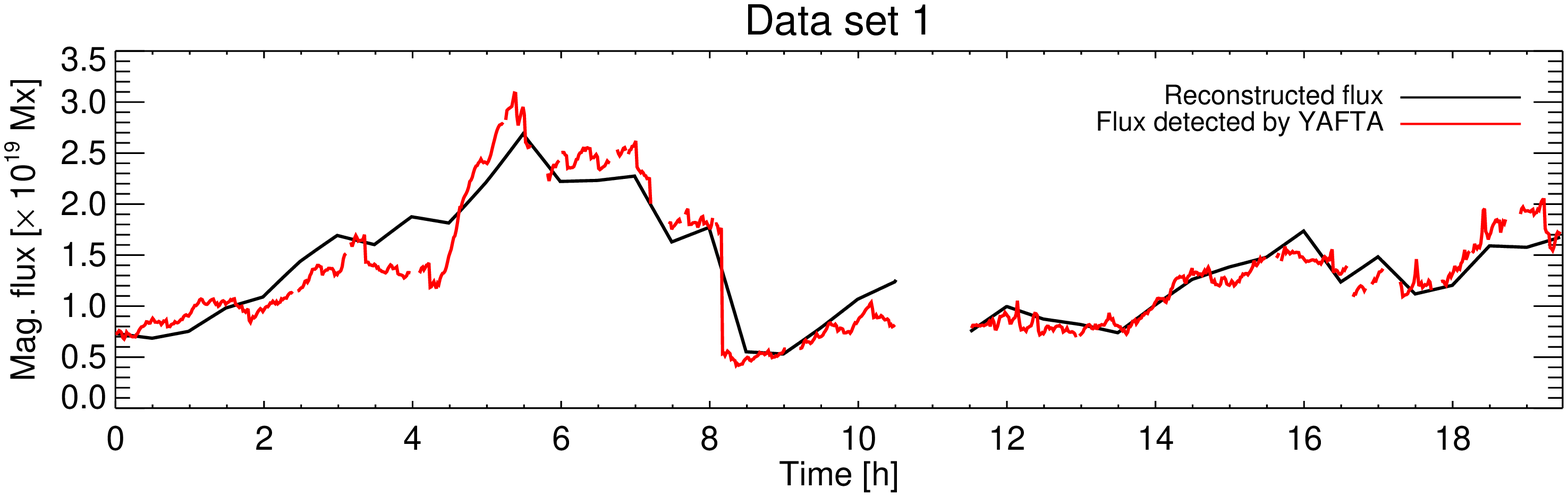}}
\resizebox{.97\hsize}{!}{\includegraphics[]{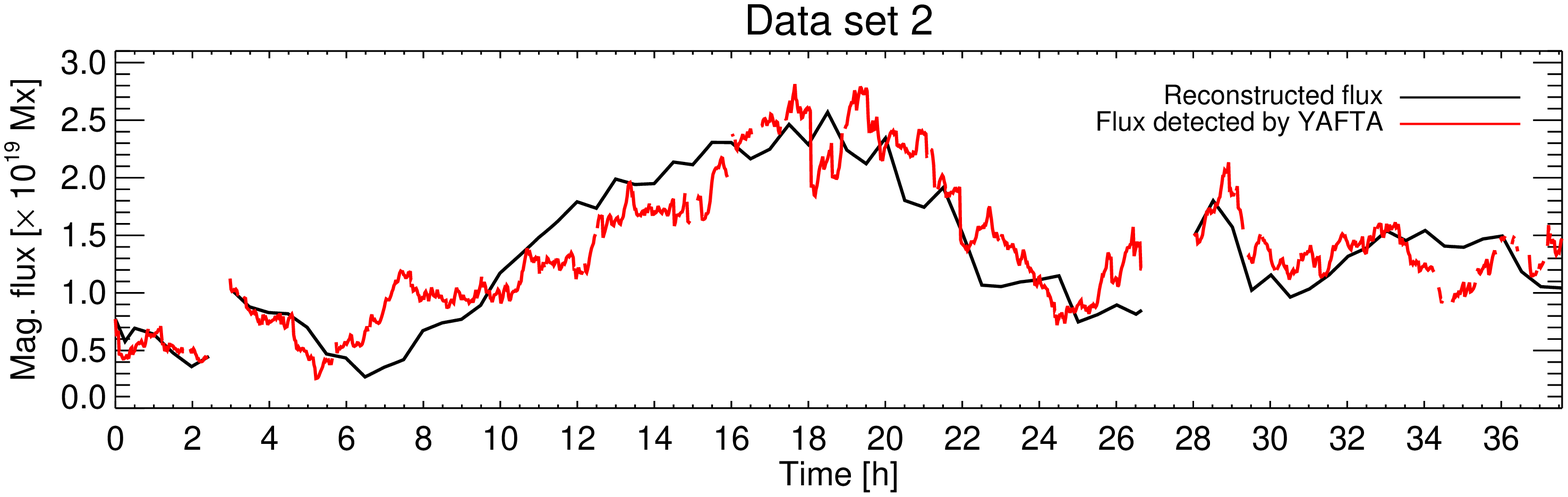}}
\end{center}
\vspace*{-1em}
\caption{Comparison between the total unsigned flux observed in the
central supergranular cells of data sets 1 (top) and 2 (bottom) and
the flux reconstructed using the instantaneous appearance and disappearance 
rates.}
\label{comparison}
\end{figure*}

\subsection{Flux appearance and disappearance rates in the IN}

Table~\ref{table1} summarizes the average appearance and disappearance
rates obtained from the analysis of the \textit{Hinode}/NFI observations, along
with their ratio. As can be seen, magnetic flux appears in the IN at a
rate of $120 \pm 3$ Mx~cm$^{-2}$~day$^{-1}$ ($40 \pm
1$~Mx~cm$^{-2}$~day$^{-1}$ if the initial flux of the magnetic
elements is considered). The total flux appearance rate is nearly the
same in the two individual supergranular cells studied here, but it
changes significantly whenever clusters of magnetic elements emerge
into the surface. At those moments, the instantaneous rates can be as
large as 200~Mx~cm$^{-2}$~day$^{-1}$.

Our total appearance rate implies that the flux brought to the entire
solar surface by IN elements is $3.7 \times 10^{24}$~Mx~day$^{-1}$ (because the IN occupies 47\% and 54\% of the FOV in data sets 1 and 2, respectively). This value is larger than the
$3\times10^{22}$ to $1.5\times10^{24}$~Mx~day$^{-1}$ injected by bipolar ephemeral regions \cite[]{Schrijver, Title, Chae, Hagenaar2001, Hagenaar2008}, the $2.6\times10^{24}$~Mx~day$^{-1}$ brought by horizontal IN fields \citep{Lites1996}, and the
$\sim$$10^{24}$~Mx~day$^{-1}$ carried by small-scale magnetic loops emerging in the solar IN \cite[]{Zirin1987, MartinezLuis}. On the
other hand, our rates are lower than the $3\times 10^{25}$ and
$3.8\times 10^{26}$~Mx~day$^{-1}$ reported by \cite{ThorntonParnell} and \cite{Zhou2013}, also based on \textit{Hinode}/NFI measurements. Still, they are enormous.

The removal of flux from supergranular cells occurs mainly through
interactions of IN features with NE patches and in-situ fading, at
rates of $50\pm3$ and $53\pm
7$~Mx~cm$^{-2}$~day$^{-1}$, respectively. These two mechanisms account
for about 40\% and 42\% of the total flux lost by IN regions. The rest 
disappears by cancellation of IN elements, at a
rate of $23\pm3$~Mx~cm$^{-2}$~day$^{-1}$ which is nearly the same in
the two supergranules we have studied.

The total disappearance rate implied by the three mechanisms capable
of removing flux from the IN is similar to the rate at which the
supergranules gain flux (see Table~\ref{table1}). They coincide to
within 7\%, reflecting the steady-state nature of the solar IN
demonstrated by Figure~5 of Paper~I. The small imbalance we observe is
probably not real, but a result of our limited magnetic sensitivity
and ability to interpret interactions between IN patches.

\subsection{Magnetic flux history of supergranular cells}

The instantaneous flux appearance and disappearance rates derived in
the previous sections allow us to reproduce the temporal evolution of
the flux in the interior of supergranular cells.
Figure~\ref{comparison} shows a comparison between the total unsigned
flux $F(t)$ observed in the supergranules of data sets 1 and 2 and the
result of integrating the instantaneous rates over time as
\begin{equation}
F(t)= F(t_0) + \int_{t_0}^{t} [F_{\rm app}(t)- F_{\rm disapp}(t)] 
\, A(t) \, {\rm d}t,
\end{equation}
where $F(t_0)$ is the total unsigned flux in the cell at the beginning
of the sequence, $F_{\rm app} (t)$ and $F_{\rm disapp}(t)$ are the
appearance and disappearance rates displayed in Figures~\ref{fig2} and
\ref{fig3}, and $A(t)$ is the area of the supergranular cell at time
$t$.

The agreement is excellent and testifies to the precision and
relevance of our results: for the first time, the evolution of IN flux
is reproduced to a remarkable degree of accuracy.

\subsection{Origin of IN flux}

\begin{figure}[t]
\begin{center}
\resizebox{.99\hsize}{!}{\includegraphics[bb=40 0 430 304]{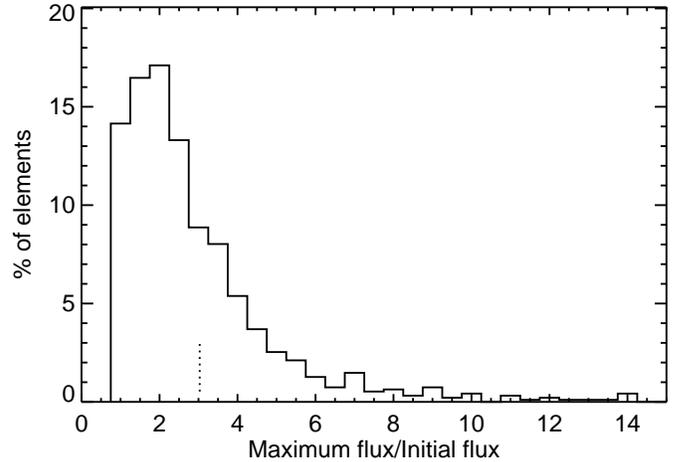}}
\end{center}
\vspace*{-1em}
\caption{Histogram of the ratio between the maximum and the initial
flux of the IN elements that appear in situ, live for at least 4 frames
(6 minutes), and never interact with other elements in data set 2. 
The mean and median of the distribution are 3.0 and 2.4, respectively.}
\label{flux_ratio}
\end{figure}

Comparing the total and initial flux appearance rates it is clear that
magnetic elements grow in flux upon appearance. Indeed, the
average initial flux of IN elements is $1.3\times 10^{16}$~Mx but, 
as can be seen in Figure~\ref{flux_ratio}, their maximum flux is on
average three times larger. This has been determined using all IN
elements that appear in situ, live for more than 6 minutes, and never
interact with other features \citep{milan_master}, so it is an
intrinsic change. Such a flux increase explains why the ratio between
the total and initial flux appearance rates is also about 3.

The fact that magnetic elements gain flux after their appearance is
puzzling but may convey important information on the origin of the IN
flux. One possibility is that the increase is simply due to continuous
flux emergence on the surface. Another possibility is that the
magnetic features are formed by coalescence of undetected background
flux which is too weak to stand out above the noise level until
sufficient flux has accumulated, as proposed by \cite{Lamb2008,
  Lamb2010}. This process may continue after the feature is first
observed, increasing its flux. Changes in the magnetic field
inclination, with the structures becoming more vertical with time, is
another possible explanation for the flux increase detected in
longitudinal magnetograms \citep{BellotOrozco}. It is necessary to
confirm whether or not these mechanisms operate on the solar surface
and understand how they affect observations, because of their
implications. For example, if a large fraction of the flux that
appears in situ is due to coalescence and not to genuine bipolar
emergence, then the actual amount of new flux brought to the surface
may be significantly smaller than suggested by current analyses. In
that case, part of the flux observed in supergranular cells would
actually be recycled (network?) flux whose magnetic connectivity is
impossible to trace after substantial reprocessing.

To clarify these issues, we will investigate the modes of appearance
and the properties of IN magnetic features---including their intrinsic
increase of flux---in the next paper of this series.

\acknowledgments 

The data used here were acquired in the framework of the \textit{Hinode}
Operation Plan 151 {\em ``Flux replacement in the solar network 
and internetwork.''} We thank the \textit{Hinode} Chief Observers for the 
efforts they made to accommodate our demanding observations. \textit{Hinode} is 
a Japanese mission developed and launched by ISAS/JAXA, with NAOJ as a
domestic partner and NASA and STFC (UK) as international partners. It
is operated by these agencies in co-operation with ESA and NSC
(Norway). MG acknowledges a JAE-Pre fellowship granted by Agencia
Estatal Consejo Superior de Investigaciones Cient\'{\i}ficas (CSIC)
toward the completion of a PhD degree. This work has been funded by
the Spanish Ministerio de Econom\'{\i}a y Competitividad through
projects AYA2012-39636-C06-05 and ESP2013-47349-C6-1-R, including a
percentage from European FEDER funds. Use of NASA's Astrophysical Data
System is gratefully acknowledged.

\end{document}